\documentclass[12pt]{article}
\usepackage{epsfig}

\begin{document}

\begin{titlepage}
\begin{flushright}
gr-qc/0506089 \\ UFIFT-QG-05-04 \\ CRETE-05-15
\end{flushright}

\vspace{0.5cm}

\begin{center}
\bf{A MEASURE OF COSMOLOGICAL ACCELERATION}
\end{center}

\vspace{0.3cm}

\begin{center}
N. C. Tsamis$^{\dagger}$
\end{center}
\begin{center}
\it{Department of Physics, University of Crete \\
GR-710 03 Heraklion, HELLAS.}
\end{center}

\vspace{0.2cm}

\begin{center}
R. P. Woodard$^{\ddagger}$
\end{center}
\begin{center}
\it{Department of Physics, University of Florida \\
Gainesville, FL 32611, UNITED STATES.}
\end{center}

\vspace{0.3cm}

\begin{center}
ABSTRACT
\end{center}
\hspace{0.3cm}
We propose quantifying the quantum gravitational back-reaction on
inflation with an invariant measure of the local acceleration rather
than the expansion rate. Our observable is suitable for models in
which there is no scalar inflaton to provide a preferred velocity
field with which to define the expansion. As an example, we use
stochastic techniques to evaluate the local acceleration at one 
loop order for $\Lambda$-driven inflation in pure quantum gravity.
\vspace{0.3cm}

\begin{flushleft}
\end{flushleft}

\vspace{0.1cm}

\begin{flushleft}
$^{\dagger}$ e-mail: tsamis@physics.uoc.gr \\
$^{\ddagger}$ e-mail: woodard@phys.ufl.edu
\end{flushleft}

\end{titlepage}

$\bullet$ {\it Introduction:} Everyone who has taught introductory physics 
is familiar with the confusion students sometimes experience between velocity 
and acceleration. When a stone is thrown upwards in a uniform gravitational 
field its acceleration is a negative constant whereas its velocity decreases 
linearly from positive to negative. If the same process were viewed from the 
frame of an inertial observer initially moving upwards faster than the stone, 
the stone's velocity would be always negative but its acceleration would be 
unchanged. We shall argue that a similar confusion exists in identifying an 
invariant measure of the quantum gravitational back-reaction on inflation. 
What one {\it wants} to do is quantify the tendency for quantum processes to
decelerate the universe, and this is independent of how rapidly the universe
may be expanding with respect to some velocity field.

Simple arguments indicate that gravitational back-reaction should slow
inflation, either in pure gravity \cite{TW1} or in scalar-driven inflation
\cite{ABM1}. The idea is that inflation rips from the vacuum a continuous
stream of virtual long wavelength quanta which are massless and not
conformally invariant. The gravitational interactions between these
particles induces a negative energy density that gradually increases as
more and more of these particles come into causal contact with one another.

Support for these ideas has come from detailed perturbative computations,
at 2-loop order in $\Lambda$-driven inflation \cite{TW2} and at 1-loop order
in scalar-driven inflation \cite{ABM2}. Unruh quite properly criticized
these results on the grounds that they were obtained by regarding the
expectation value of the gauge-fixed metric as an observable \cite{Unruh}.
One way of resolving his concern is to do the computation in other gauges.
That was too difficult for the 2-loop gravitational process but a simple
check in a completely different gauge showed no change in the claimed
back-reaction for scalar-driven inflation \cite{AW1}.

A much better way of addressing Unruh's criticism is to construct an
invariant {\it operator} that measures the expansion rate even when
perturbations are present. One can then compute the expectation value of
this operator. A crude scalar measure of the spacetime expansion rate
was obtained using the inverse conformal d`Alembertian acting upon unity
\cite{AW2}. This operator can be promoted to an invariant be evaluating
it at an physically defined observation point. When the 1-loop expectation 
value of that invariant was evaluated for a model which had previously 
seemed to show back-reaction at one loop, the result was zero secular 
back-reaction \cite{AW3}. Curiously, the scalar observable contributed 
nothing to this. When its expectation value was computed in the old, 
gauge-dependent coordinate systems it continued to show precisely the 
old back-reaction effect. The nullification of the effect was entirely
due to the corrections needed to define the observation point invariantly.

A major step forward was taken by Geshnizjani and Brandenberger \cite{GB1},
who worked out how to apply the standard measure of expansion \cite{SWH}
to the case of scalar-driven inflation. In this situation the scalar
inflaton $\varphi(t,\vec{x})$ provides a preferred 4-velocity field:
\footnote{Hellenic indices take on spacetime values while Roman indices
take on space values. Our conventions are that the metric $g_{\mu\nu}$
has spacelike signature and the curvature tensor equals
$R^{\alpha}_{~\beta \mu \nu} \equiv
\Gamma^{\alpha}_{~\nu \beta , \mu} + 
\Gamma^{\alpha}_{~\mu \rho} \; \Gamma^{\rho}_{~ \nu \beta} 
\, - \, (\mu \leftrightarrow \nu)$.}

\begin{equation}
u^{\mu}(t,\vec{x}) \, \equiv \,
- \frac{g^{\mu\nu}(t,\vec{x}) \; 
\partial_{\nu} \varphi(t,\vec{x})}
{\sqrt{-g^{\alpha\beta}(t,\vec{x}) \; 
\partial_{\alpha} \varphi(t,\vec{x}) \;
\partial_{\beta} \varphi(t,\vec{x})}} 
\;\; . \label{u}
\end{equation}
In this expression we mean the quantum metric and the full inflaton operator,
including its classical background part and the quantum perturbation:
\begin{equation}
\varphi(t,\vec{x}) \, = \, \varphi_0(t) + \delta\varphi(t,\vec{x}) 
\;\; .
\end{equation}
Expression (\ref{u}) for $u^{\mu}(t,\vec{x})$ obviously gives a timelike unit
vector which is well-defined for small perturbations. By taking one third of
its divergence one obtains a scalar measure of the expansion rate. The scalar
becomes a full invariant when evaluated at a physically defined observation
point. Owing to the spatial homogeneity and isotropy of the state it is only
necessary to invariantly fix the time. The preferred coordinate system is one
in which the full scalar agrees with its background value. That is, one solves 
perturbatively for the operator $\tau[\varphi,g](t,\vec{x})$ which enforces 
the condition:
\begin{equation}
\varphi\Bigl(\tau(t,\vec{x}),\vec{x}\Bigr) \, = \, \varphi_0(t) 
\;\; . \label{tau}
\end{equation}
The invariant expansion measure is:
\begin{equation}
\mathcal{H}(t,\vec{x}) \; \equiv \;
\frac13 \, u^{\mu}_{~ ;\mu} \Bigl( \tau(t, \vec{x}), \vec{x} \Bigr)
\, = \, \frac13 \, 
\frac{1}{\sqrt{-g}} \, \partial_{\mu} \left( \sqrt{-g} \, u^{\mu} \right)
\Bigl( \tau(t, \vec{x}), \vec{x} \Bigr)
\;\; . \label{exp}
\end{equation}

When the expectation value of (\ref{exp}) was evaluated for models which
seemed to show secular back-reaction at 1-loop order, the result was no 
secular effect \cite{GB1}. Computations of quantities such as the stress 
tensor \cite{FMVV1} continue to show secular back-reaction at one loop when 
observables are evaluated at a gauge-fixed coordinate point instead of at 
the physically defined location (\ref{tau}). It has also been shown that 
secular back-reaction can seem to occur at 1-loop order in a two-scalar 
model if one uses the other, spectator, scalar to measure the expansion 
rate \cite{GB2}.

It seems clear that there is no secular one loop back-reaction in 
conventional scalar-driven inflation \cite{MB}.
\footnote{But unconventional models might show it \cite{GA,BM}.} 
This actually agrees with the putative physics of the effect. The causative 
agent is 1-loop particle production so it would have been fortuitous to see 
a secular gravitational response at the same order. Secular growth can only 
come as more and more of the produced particles come into interaction with 
one another and that must be delayed -- in perturbation theory -- until at 
least 2-loop order. Scalar models can be contrived which do show an 
invariantly quantified back-reaction effect at higher loop order \cite{AW4}. 
And the problem will not go away because of the enormous potential impact 
of back-reaction in a realistic model \cite{RHB1,Ford1,RHB2,BMz}.

A particularly interesting model in which to investigate back-reaction
is $\Lambda$-driven inflation. The Lagrangian is just that of gravity
with a positive cosmological constant:
\begin{equation}
\mathcal{L} \; = \; \frac1{16 \pi G} \,
\Bigl( R - 2 \Lambda \Bigr) \sqrt{-g} 
\;\; , \label{Lgr}
\end{equation}
and the perturbative background would be de Sitter. One must certainly go 
to 2-loop order to see back-reaction in this model \cite{Ford2,FMVV2,TW3}. 
On the other hand, the system is uncomplicated by the classical evolution 
of a scalar inflaton. The classical result for this model is no evolution 
at all so, if one sees evolution then it is due to quantum gravitational 
back-reaction; {\it except} that one still has to invariantly quantify the 
effect and there is no scalar inflaton to serve as a clock everyone can 
agree upon. Of course one could simply define an ersatz scalar any number 
of ways. For example, one could take the invariant volume of the past 
light-cone, measured from the observation point back to the initial value 
surface upon which the quantum state is released. This transforms as a 
scalar, and it certainly increases monotonically in the timelike direction. 
The trouble is that many such ``scalars'' can be defined and none of them 
has the privileged role that the scalar inflaton plays in setting the zero 
of time for the post-inflationary universe. \\

$\bullet$ {\it The de Sitter example:} The ambiguity and the conflicting
results it can produce are easy to understand on the classical level in
de Sitter spacetime. Suppose we take the ``scalar'' to be the time $t_c$
of the ``closed'' co-ordinate system:
\footnote{De Sitter spacetime has the topology of $S^{D-1} \times \Re$ 
and it is natural to cover the full manifold by using a co-ordinate 
system in which the spatial sections are $S^{D-1}$.}
\begin{equation}
Closed
\qquad \Longrightarrow \qquad
ds^2 \; = \;
- \, dt_c^2 \, + \, \frac{1}{H^2} \, \cosh^2(H t_c) \; d\Omega^2_3 
\;\; , \label{closedds2} 
\end{equation}
\begin{equation}
-\infty < t_c < \infty 
\quad , \quad
0 \leq \alpha_i \leq \pi \;\; (i = 1 , 2)
\quad , \quad
0 \leq \alpha_3 < 2\pi
\;\; . \label{closedcoord}
\end{equation}
It is elementary to compute the velocity field (\ref{u}) and the expansion
rate (\ref{exp}) in this system:
\begin{equation}
Closed
\qquad \Longrightarrow \qquad
u^{\mu}_c \, = \, \delta^{\mu}_{~ 0}
\quad \Longrightarrow \quad
H_c \, = \, H \tanh (Ht_c)
\;\; . \label{closedvel}
\end{equation}
The integral curves of this velocity field are initially drawing together 
-- so the expansion rate starts negative -- but they eventually draw apart, 
resulting in positive expansion.

Now suppose we adopt as our ``scalar'' the time $t_o$ of the ``open'' 
co-ordinates sub-manifold:
\footnote{The ``open'' co-ordinate system covers half of the full de 
Sitter manifold and has the topology of $\Re^D$.}  
\begin{equation}
Open 
\qquad \Longrightarrow \qquad
ds^2 \; = \;
- \, dt_o^2 \, + \, e^{2Ht_o} \, \Vert d{\vec x} \Vert^2
\;\; . \label{opends2}
\end{equation}
\begin{equation}
-\infty < t_o < \infty
\quad , \quad
0 \leq \Vert {\vec x} \Vert < \infty
\;\; \label{opencoord}
\end{equation}
The resulting velocity field (\ref{u}) and expansion rate (\ref{exp}) are:
\begin{equation}
Open 
\qquad \Longrightarrow \qquad
u^{\mu}_o  \, = \, \delta^{\mu}_{~ 0}
\quad \Longrightarrow \quad
H_o \, = \, H
\;\; . \label{openvel}
\end{equation}
For the ``open'' velocity field the expansion rate is a positive constant 
throughout the sub-manifold. The latter includes some of the very same 
points for which the expansion rate is {\it negative} when measured with 
the velocity field $u^{\mu}_c$. When this kind of ambiguity exists on the 
classical background there is little hope of finding a measure of spacetime
expansion upon which all observers can agree when quantum perturbations
are present. 

A little thought leads to the realization that -- in attempting to 
identify a preferred velocity field from which to measure expansion 
in $\Lambda$-driven inflation -- we have succumbed to the same confusion
as the introductory physics student who seeks a preferred inertial frame.
In the latter case, the laws of physics are phrased in terms of
acceleration. In Galilean relativity, the co-ordinates of a given point 
in two different inertial frames of reference -- which are in relative
motion with velocity ${\vec V}$ with respect to one another -- are related 
by:
\begin{eqnarray}
{\vec r} & = &
{\vec r^{\hspace{0.05cm} \prime}} \; + \; {\vec V} t
\;\; , \label{galilean1} \\
t & = & t^{\prime}
\;\; . \label{galilean2} 
\end{eqnarray}
The position and velocity at a given point are frame-dependent quantities
while the acceleration is not.

So too in quantum gravity, it is really the local cosmological 
{\it acceleration} we wish to measure and not the expansion in some
velocity field. \\

$\bullet$ {\it Acceleration in de Sitter:}  In general, the acceleration 
is obtained from the deviation equation of two infinitesimally close 
geodesics $\chi^{\mu}(\tau)$ and $\chi^{\mu}(\tau) + \delta\chi^{\mu}(\tau)$:
\begin{equation}
\frac{D^2 \, \delta\chi^{\mu}(\tau)}{D \tau^2} \; = \;
- R^{\mu}_{~ \nu \rho \sigma} \left[ \chi(\tau) \right] \; 
{\dot \chi}^{\nu}(\tau) \;
\delta\chi^{\rho}(\tau) \; 
{\dot \chi}^{\sigma}(\tau)
\;\; . \label{geodev}
\end{equation}
The curvature tensor for de Sitter spacetime equals:
\begin{equation}
de \; Sitter
\qquad \Longrightarrow \qquad
R^{\alpha}_{~ \beta \gamma \delta} \; = \;
H^2 \left( \delta^{\alpha}_{~ \gamma} \; g_{\beta \delta} 
\, - \, \delta^{\alpha}_{~ \delta} \; g_{\beta \gamma} \right) 
\;\; . \label{dScurv}
\end{equation}
We now consider two initially parallel, timelike geodesics with 
spacelike separation $\Delta^{\mu}$. In synchronous gauge we
have:
\begin{eqnarray}
\chi^{\mu}(\tau) & = & \tau \, \delta^{\mu}_{~0} 
\;\; , \label{timegeo1} \\
\chi^{\mu}(\tau) \, + \, \delta\chi^{\mu}(\tau) & = & 
\tau \, \delta^{\mu}_{~0} \, + \, \Delta^{\mu}(0)
\;\; , \label{timegeo2} \\
g_{\mu\nu} \; {\dot \chi}^{\mu} \; \Delta^{\nu} \; = \; 0
& , &
g_{\mu\nu} \; {\dot \chi}^{\mu} \; {\dot \chi}^{\nu} \; = \; -1
\;\; , \label{paralleltimelike}
\end{eqnarray}
so that:
\begin{equation}
\frac{d\chi^{\mu}(\tau)}{d\tau} \, \equiv \,
{\dot \chi^{\mu}} \, = \, 
\delta^{\mu}_{~ 0}
\;\; . \label{derivative}
\end{equation}
For these geodesics, the deviation equation (\ref{geodev}) takes
the form:
\begin{eqnarray}
\frac{D^2 \, \Delta^{\mu}(\tau)}{D \tau^2} 
& = &
- H^2 \Delta^{\mu} \; g_{\rho \sigma} \; 
{\dot \chi}^{\rho} \; {\dot \chi}^{\sigma} \; + \; 
H^2 {\dot \chi}^{\mu} \; g_{\rho \sigma} \; 
{\dot \chi}^{\rho} \; \Delta^{\sigma} 
\;\; , \label{dSgeodev1} \\
& = &
H^2 \, \Delta^{\mu}
\;\; , \label{dSgeodev2}
\end{eqnarray}
and a constant positive acceleration is manifest for any point on the 
manifold: the deviation between initially parallel, timelike and freely 
falling observers expands exponentially at all points in de Sitter 
spacetime. This is a frame invariant statement that characterizes the 
de Sitter geometry and, consequently, is free of the ambiguities that 
plague the velocity field. \\

$\bullet$ {\it Acceleration in general:} It is convenient to use the
freedom under general co-ordinate transformations to bring an arbitrary 
metric into synchronous gauge:  
\begin{equation}
ds^2 \; = \;
- \, dt^2 \, + \, g_{ij}(t, {\vec x}) \, dx^i \, dx^j
\;\; . \label{ds2}
\end{equation}
The geodesic deviation equation reduces to::
\begin{eqnarray}
\frac{D^2 \, \Delta^i(\tau)}{D \tau^2} 
& = &
- \, R^i_{~ 0j0} \; \Delta^j
\;\; , \label{reduction} \\
& = &
- \left( - \, \frac12 \, g^{ik} \, {\ddot g}_{kj} \; + \; 
\frac14 \, g^{ik} \, g^{lm} \,
{\dot g}_{kl} \, {\dot g}_{mj} \right) \Delta^j
\;\; , \label{rhs}
\end{eqnarray}
The cosmological observation we are interested should not depend on 
the direction of the vector $\Delta^i$, hence we contract into the 
vector by multiplying with $g_{ij} \, \Delta^j$. Nor on the magnitude 
of the vector $\Delta^i$, hence we scale the vector by dividing with 
its magnitude $g_{rs} \, \Delta^r \, \Delta^s$:
\begin{equation}
\frac{g_{ij} \, \Delta^j}{g_{rs} \, \Delta^r \, \Delta^s} \,
\frac{D^2 \, \Delta^i(\tau)}{D \tau^2} \; = \;
\frac{1}{g_{rs} \, \Delta^r \, \Delta^s}
\left( \, \frac12 \, {\ddot g}_{ij} \; - \; 
\frac14 \, g^{kl} \, {\dot g}_{ik} \, 
{\dot g}_{jl} \right) \Delta^i \, \Delta^j
\;\; , \label{contractdivide}
\end{equation} 
It is the right hand side of the geodesic deviation equation 
(\ref{geodev}) which should provide the cosmological acceleration 
measurement $\gamma$ that the observer performs at event $x$:
\begin{equation}
\gamma (x) \; \equiv \;
\frac{1}{g_{rs} \, \Delta^r \, \Delta^s}
\left( \, \frac12 \, {\ddot g}_{ij} \; - \; 
\frac14 \, g^{kl} \, {\dot g}_{ik} \, 
{\dot g}_{jl} \right) \Delta^i \, \Delta^j
\;\; . \label{cosmoacc}
\end{equation}

$\bullet$ {\it Acceleration for flat Robertson-Walker spacetimes:}
Let us restrict ourselves to the cosmologically interesting homogeneous, 
isotropic and spatially flat geometries:
\begin{equation}
FRW
\qquad \Longrightarrow \qquad
g_{ij} (t, {\vec x}) \; \equiv \; a^2 (t) \, \delta_{ij}
\;\; . \label{FRW}
\end{equation}
Derivatives of the scale factor $a(t)$ give the velocity (Hubble) 
parameter $H(t)$ and the deceleration parameter $q(t)$:
\begin{equation}
H(t) \; \equiv \; \frac{\dot{a}}{a}
\qquad , \qquad
q(t) \; \equiv \; -\frac{a \ddot{a}}{\dot{a}^2}
\, = \, -1 - \frac{\dot{H}}{H^2}
\;\; . \label{H&q}
\end{equation}
The relevant components of the curvature tensor are:
\begin{equation}
FRW
\qquad \Longrightarrow \qquad
R^i_{~ 0j0} \, = \,
- \left( H^2 \, + \, {\dot H} \right) \, \delta^i_{~ j} \, = \, 
q H^2 \, \delta^i_{~ j}
\;\; , \label{Ri0j0}
\end{equation}
and, therefore, the cosmological acceleration observable $\gamma$
equals:
\begin{equation}
FRW
\qquad \Longrightarrow \qquad
\gamma (t) \; = \; - \, q H^2 \; = \;
\frac{\ddot a}{a}
\;\; , \label{FRWcosmoacc}
\end{equation}
and it measures the {\it fractional} local cosmological acceleration. \\

$\bullet$ {\it Stochastic Acceleration:} It is simple to evaluate the
expectation value of the local acceleration $\gamma$ at one-loop order
for the gravitational action (\ref{Lgr}) using stochastic techniques
\cite{starobinsky1,starobinsky2,TW4} under the assumption of no 
back-reaction. First, we conformally re-scale the metric and express 
its time derivatives as functions of the re-scaled metric 
${\widetilde g}_{ij}$:
\begin{eqnarray}
g_{ij} \equiv a^2 \, {\widetilde g}_{ij}
& \Longrightarrow &
{\dot g}_{ij} \; = \;
2Ha^2 \, {\widetilde g}_{ij} \, + \,
a^2 \, {\dot {\widetilde g}}_{ij}
\;\; , \label{gtilde} \\
& \Longrightarrow &
{\ddot g}_{ij} \; = \;
\left( 2{\dot H} \, + \, 4H^2 \right) a^2 \, {\widetilde g}_{ij} \, + \,
4H a^2 \, {\dot {\widetilde g}}_{ij} \, + \, 
a^2 \, {\ddot {\widetilde g}}_{ij}
\;\; . \nonumber 
\end{eqnarray}
We also define the Euclidean direction vector:
\begin{equation}
n^i \equiv \frac{\Delta^i}{\sqrt{\Delta^j \Delta^j}}
\;\; . \label{unitdirection} 
\end{equation}
Our observable (\ref{cosmoacc}) becomes:
\begin{equation}
\gamma (x) \; = \;
H^2 \, + \, {\dot H} \, + \, 
\frac{1}{{\widetilde g}_{rs} \, n^r n^s}
\left( \, \frac12 \, {\ddot {\widetilde g}}_{ij} \, + \, 
H \, {\dot {\widetilde g}}_{ij} \, - \, 
\frac14 \, {\widetilde g}^{kl} \, {\dot {\widetilde g}}_{ik} \, 
{\dot {\widetilde g}}_{jl} \right) n^i n^j
\;\; . \label{cosmoacc2}
\end{equation}
Under the assumption of no back-reaction, the re-scaled spatial metric
consists of the identity element plus just the transverse traceless 
field $h_{ij}^{TT}$ of linearized gravitons:
\begin{equation}
No \; Back \!-\! Reaction
\qquad \Longrightarrow \qquad
{\widetilde g}_{ij} \; = \;
\delta_{ij} \, + \, \sqrt{32\pi G} \, h_{ij}^{TT}
\;\; , \label{nonback}
\end{equation}
and to linear order in $G$, the local acceleration operator is:
\begin{eqnarray}
\gamma (x) \!\! & = & \!\!
H^2 \, + \, {\dot H} \, + \, \Biggl[ \,
\frac12 \sqrt{32\pi G} \, {\ddot h}_{ij}^{TT} \, + \, 
H \sqrt{32\pi G} \, {\dot h}_{ij}^{TT} \, - \, 
8 \pi G \, {\dot h}_{ik}^{TT} \, {\dot h}_{jk}^{TT} 
\nonumber \\
&\mbox{} &
- \left( \,
16\pi G \, {\ddot h}_{ij}^{TT} \, + \, 
32\pi G H {\dot h}_{ij}^{TT} 
\, \right) h_{kl}^{TT} \, n^k n^l 
\, \Biggr] n^i n^j \; + \; O(G^{\frac32})
\;\; , \label{nobackaccel}
\end{eqnarray}
Taking the average over all directions and using:
\begin{equation}
n^i n^j \; \longrightarrow \; \frac13 \, \delta^{ij}
\quad , \quad
n^i n^j n^k n^l \; \longrightarrow \; 
\frac{1}{15} \left( 
\delta^{ij} \, \delta^{kl} \, + \,
\delta^{ik} \, \delta^{jl} \, + \,
\delta^{il} \, \delta^{jk} \right)
\;\; . \label{angave}
\end{equation}
results in the following expression:
\begin{eqnarray}
\int d^2n \; \gamma (x) \!\! & = & \!\!
H^2 \, + \, {\dot H} \, - \, 
\frac{8\pi G}{3} \, {\dot h}_{ij}^{TT} \, {\dot h}_{ij}^{TT} 
\, - \, \frac{32\pi G}{15} \, {\ddot h}_{ij}^{TT} \, h_{ij}^{TT}
\nonumber \\
& \mbox{} &
- \, \frac{64\pi GH}{15} \, {\dot h}_{ij}^{TT} \, h_{ij}^{TT} 
\; + \; O(G^{\frac32})
\;\; . \label{nobackaccelpert}
\end{eqnarray}
Before substituting the stochastic mode expansion for $h_{ij}^{TT}$ in 
(\ref{nobackaccelpert}) we must drop the terms where two derivatives act 
on the same field or upon different fields 
\cite{starobinsky1,starobinsky2,TW4}:
\begin{equation}
\mbox{}
\hspace{-0.3cm}
Stochastic
\quad \Longrightarrow \quad
\int d^2n \; \gamma (x) \, = \,
H^2 \, + \, {\dot H} \, - \, 
\frac{64\pi GH}{15} \, {\dot h}_{ij}^{TT} \, h_{ij}^{TT} 
\, + \, O(G^{\frac32})
\label{stochaccelpert}
\end{equation}

Now from the expansion of the stochastic field in de Sitter spacetime
\cite{starobinsky1,starobinsky2,TW4}:
\footnote{In (\ref{stochfield}), $c.c.$ denotes the conjugate expression.} 
\begin{equation}
h_{ij}^{TT} (t, {\vec x}) \, = \,
\int \frac{d^3k}{(2\pi)^3} \;
\theta(H a(t) - k) 
\sum_{\lambda} \Biggl\{
\frac{H}{\sqrt{2k^3}} \, e^{i {\vec k} \cdot {\vec x}}
\epsilon_{ij}({\vec k}, \lambda) \; \alpha_{{\vec k}, \, \lambda}
\; + \; c.c. \Biggr\}
\label{stochfield}
\end{equation}
it is elementary to calculate its two-point function:
\begin{eqnarray}
\Big\langle \; 
h_{ij}^{TT}(t,\vec{x}) \; h_{ij}^{TT}(t',\vec{x}) 
\; \Big\rangle
\!\! & = & \!\!
\int \frac{d^3k}{(2\pi)^3} \;
\theta(H a(t) - k) \,\, \theta(H a(t') - k) 
\nonumber \\
& \mbox{} &
\hspace{1.5cm}
\times \,\, \frac{H^2}{2k^3} \sum_{\lambda} 
\epsilon_{ij}({\vec k}, \lambda) \; 
\epsilon_{kl}({\vec k}, \lambda) 
\;\; , \nonumber \\
& \mbox{} &
\hspace{-2.5cm}
= \;
\frac{2H^2}{16\pi^3} \, 4\pi
\int_H^{\infty} \frac{dk}{k} \;
\theta(H a(t) - k) \,\, \theta(H a(t') - k) 
\;\; , \nonumber \\
& \mbox{} &
\hspace{-2.5cm}
= \;
\frac{H^2}{2\pi^2} \Bigl\{ \,
\theta(t - t') \, \ln \left[ a(t') \right] \, + \,
\theta(t' - t) \, \ln \left[ a(t) \right] 
\, \Bigr\}
\;\; , \nonumber \\
& \mbox{} &
\hspace{-2.5cm}
= \;
\frac{H^3}{2\pi^2} \Bigl[ \,
\theta(t - t') \; t' \; + \;
\theta(t' - t) \; t \, \Bigr]
\;\; . \label{stochfield2pt}
\end{eqnarray}
The final answer emerges when we use (\ref{stochfield2pt}) in expression 
(\ref{stochaccelpert}):
\begin{equation}
de \; Sitter
\quad \Longrightarrow \quad
\Big\langle \, 
\int d^2n \; \gamma (x)
\, \Big\rangle 
\, = \,
H^2 \left[ \, 1 \, - \, \frac{32}{15\pi} \, G H^2
\, + \, O(G^2) \, \right]
\; . \label{stochaccelresult}
\end{equation}
Even for GUT-scale inflation the correction is a very small {\it constant}
effect. Because the stochastic formalism correctly captures only the
leading infrared logarithms \cite{starobinsky1,starobinsky2,TW4}, our 
result (\ref{stochaccelresult}) is consistent with zero change at one-loop
order which is, in turn, consistent with the physics of back-reaction in
quantum gravity on de Sitter spacetime. 

\vspace{1cm}

\centerline{\bf Acknowledgements}
It is a pleasure to acknowledge years of friendly and stimulating
discussions on this topic with L. R. Abramo, R. H. Brandenberger, G.
Geshnizjani, A. Guth, D. N. Page and W. G. Unruh. This work was partially
supported by the European Social fund and National resources 
Y$\Pi$E$\Pi\Theta$-PythagorasII-2103, by European Union grants FP-6-012679 
and MRTN-CT-2004-512194, by NSF grant PHY-0244714, and by the Institute 
for Fundamental Theory at the University of Florida.


\newpage

\begin{thebibliography}{99}

\bibitem{TW1} N. C. Tsamis and R. P. Woodard, 
              Phys. Lett. {\bf B301} (1993) 351.

\bibitem{ABM1} V. F. Mukhanov, L. R. Abramo and R. H. Brandenberger,
               Phys. Rev. Lett. {\bf 78} (1997) 1624, 
               {\bf arXiv:}gr-qc/9609026.

\bibitem{TW2} N. C. Tsamis and R. P. Woodard, 
              Ann. Phys. {\bf 253} (1997) 1, \\
              {\bf arXiv:}hep-ph/9602316.

\bibitem{ABM2} L. R. Abramo, R. H. Brandenberger and V. F. Mukhanov,
               Phys. Rev. {\bf D56} (1997) 3248, 
               {\bf arXiv:}gr-qc/9704037.

\bibitem{Unruh} W. Unruh, ``Cosmological long wavelength perturbations,'' \\
                {\bf arXiv:}astro-ph/9802323.

\bibitem{AW1} L. R. Abramo and R. P. Woodard, 
              Phys. Rev. {\bf D60} (1999) 044010, 
              {\bf arXiv:}astro-ph/9811430.

\bibitem{AW2} L. R. Abramo and R. P. Woodard, 
              Phys. Rev. {\bf D65} (2002) 043507, \\
              {\bf arXiv:}astro-ph/0109271.

\bibitem{AW3} L. R. Abramo and R. P. Woodard, 
              Phys. Rev. {\bf D65} (2002) 063515, \\
              {\bf arXiv:}astro-ph/0109272.

\bibitem{GB1} G. Geshnizjani and R. H. Brandenberger, 
              Phys. Rev. {\bf D66} (2002) 123507, 
              {\bf arXiv:}gr-qc/0204074.

\bibitem{SWH} S. W. Hawking and G. F. R. Ellis, 
              {\it The Large Scale Structure of Spacetime} 
              (Cambridge University Press, Cambridge, 1975).

\bibitem{FMVV1} F. Finelli, G. Marozzi, G. P. Vacca and G. Venturi,
                Phys. Rev. {\bf D69} (2004) 123508,  
                {\bf arXiv:}gr-qc/0310086.

\bibitem{GB2} G. Geshnizjani and R. H. Brandenberger, 
              JCAP {\bf 0504} (2005) 006, \\ 
              {\bf arXiv:}hep-ph/0310265.

\bibitem{MB} P. Martineau and R. H. Brandenberger, 
             ``The Effects of Gravitational Back-Reaction of Cosmological 
               Perturbations,'' \\
             {\bf arXiv:}astro-ph/0505236.

\bibitem{GA} G. Geshnizjani and N. Afshordi, 
             JCAP {\bf 0501} (2005) 011, \\
             {\bf arXiv:}gr-qc/0405117.

\bibitem{BM} R. H. Brandenberger and J. Martin, 
             Phys. Rev. {\bf D71} (2005) 023504, \\
             {\bf arXiv:}hep-th/0410223.

\bibitem{AW4} L. R. Abramo and R. P. Woodard, 
              Phys. Rev. {\bf D65} (2002) 063516, \\
              {\bf arXiv:}astro-ph/0109273.

\bibitem{RHB1} R. H. Brandenberger, 
               ``Back-Reaction of Cosmological Perturbations,'' 
               {\bf arXiv:}hep-th/0004016.

\bibitem{Ford1} L. H. Ford, 
                ``What does Quantum Field Theory in Curved Spacetime Have 
                  to Say about the Dark Energy?'', 
                {\bf arXiv:}gr-qc/0210096.

\bibitem{RHB2} R. H. Brandenberger, 
               ``Back-Reaction of Cosmological Perturbations and the 
                 Cosmological Constant Problem,'' 
               {\bf arXiv:}hep-th/0210165.

\bibitem{BMz} R. H. Brandenberger and A. Mazumdar, 
              JCAP {\bf 0408} (2004) 015, \\
              {\bf arXiv:}hep-th/0402205.

\bibitem{Ford2} L. H. Ford, 
                Phys. Rev. {\bf D35} (1985) 710.

\bibitem{FMVV2} F. Finelli, G. Marozzi, G. P. Vacca and G. Venturi,
                Phys. Rev. {\bf D71} (2005) 023522, 
                {\bf arXiv:}gr-qc/0407101.

\bibitem{TW3} N. C. Tsamis and R. P. Woodard, 
              ``Dimensionally Regulated Graviton 1-Point Function in 
                de Sitter,'' 
              {\bf arXiv:}gr-qc/0506056.

\bibitem{starobinsky1} A. A. Starobinski\u{\i},
        ``Stochastic de Sitter (inflationary) stage in the early universe,''
        in {\it Field Theory, Quantum Gravity and Strings},
        ed. H. J. de Vega and N. Sanchez (Springer-Verlag, Berlin, 1986)
        pp. 107-126.

\bibitem{starobinsky2} A. A. Starobinski\u{\i} and J. Yokoyama,
                       Phys. Rev. {\bf D50} (1994) 6357, \\
                       {\bf arXiv:}astro-ph/9407016.

\bibitem{TW4} N. C. Tsamis and R. P. Woodard, 
              ``Stochastic Quantum Gravitational Inflation,'' 
              {\bf arXiv:}gr-qc/0505115.


\end{thebibliography}
\end{document}